\documentclass[aps,prl,reprint,showpacs,floatfix]{revtex4-1}
\pdfoutput=1

\usepackage[pdftex]{graphicx}
\usepackage{amsmath}
\usepackage{amsfonts}
\usepackage{mathrsfs}
\usepackage{bm}

\begin{document}

\title{Topological Transitions of Gapless Paired States in Mixed-Geometry Lattices}

\author{Dong-Hee Kim}
\author{Joel S. J. Lehikoinen}
\author{P\"{a}ivi T\"{o}rm\"{a}}
\email{paivi.torma@aalto.fi}

\affiliation{COMP Centre of Excellence, Department of Applied Physics, Aalto University, FI-00076 Aalto, Finland}

\begin{abstract}
We propose a mixed-geometry system of fermionic species selectively confined 
in lattices of different geometry. We investigate how such asymmetry can lead 
to exotic multiband fermion pairing in an example system of honeycomb and triangular lattices.
A rich phase diagram of interband pairing with
gapped and gapless excitations is found at zero temperature.
We find that the two-band contribution of the honeycomb lattices to the paired state 
helps to stabilize the gapless phase with one or two Fermi surfaces. 
We also show that the Fermi surface topology 
further divides the gapless phase into subclasses between which 
the system undergoes density-driven Lifshitz transitions.   
\end{abstract}

\pacs{67.85.-d,74.20.-z,03.75.Ss,71.10.Fd}

\maketitle

Fermion pairing may coexist with magnetism in unconventional superconductors 
and nuclear matter, stabilized by non-BCS mechanisms. 
The standard BCS theory considers a perfectly symmetric pairing in up- and down-spin
particles with opposite momenta lying on identical Fermi surfaces. 
One of the fundamental issues in condensed matter physics is what would happen 
if this setting were broken. For instance, two remarkable scenarios were proposed 
for pairing with mismatched Fermi surfaces between the spin components. 
The Fulde-Ferrell-Larkin-Ovchinnikov state suggests Cooper pairs 
carrying finite center-of-mass momentum~\cite{FF,LO}. 
The Sarma or breached-pair (BP) state describes the polarized superfluid of 
zero-momentum pairs separated from unpaired excess particles 
forming a Fermi sea~\cite{Sarma,BP}. 
These scenarios remain elusive yet have inspired experiments 
in solid-state materials~\cite{Radovan2003,Kenzelmann2008} 
and ultracold Fermi gases~\cite{MIT,RICE,ENS,Liao2010}.
Here, we propose another fundamental way of distorting the symmetry 
between the spin species in lattice degrees of freedom 
with which we introduce a minimal multiband setting. 
We find that gapless paired states and transitions with changing Fermi surface topology
are stabilized with multiband effects.

\begin{figure}
\includegraphics[width=0.48\textwidth]{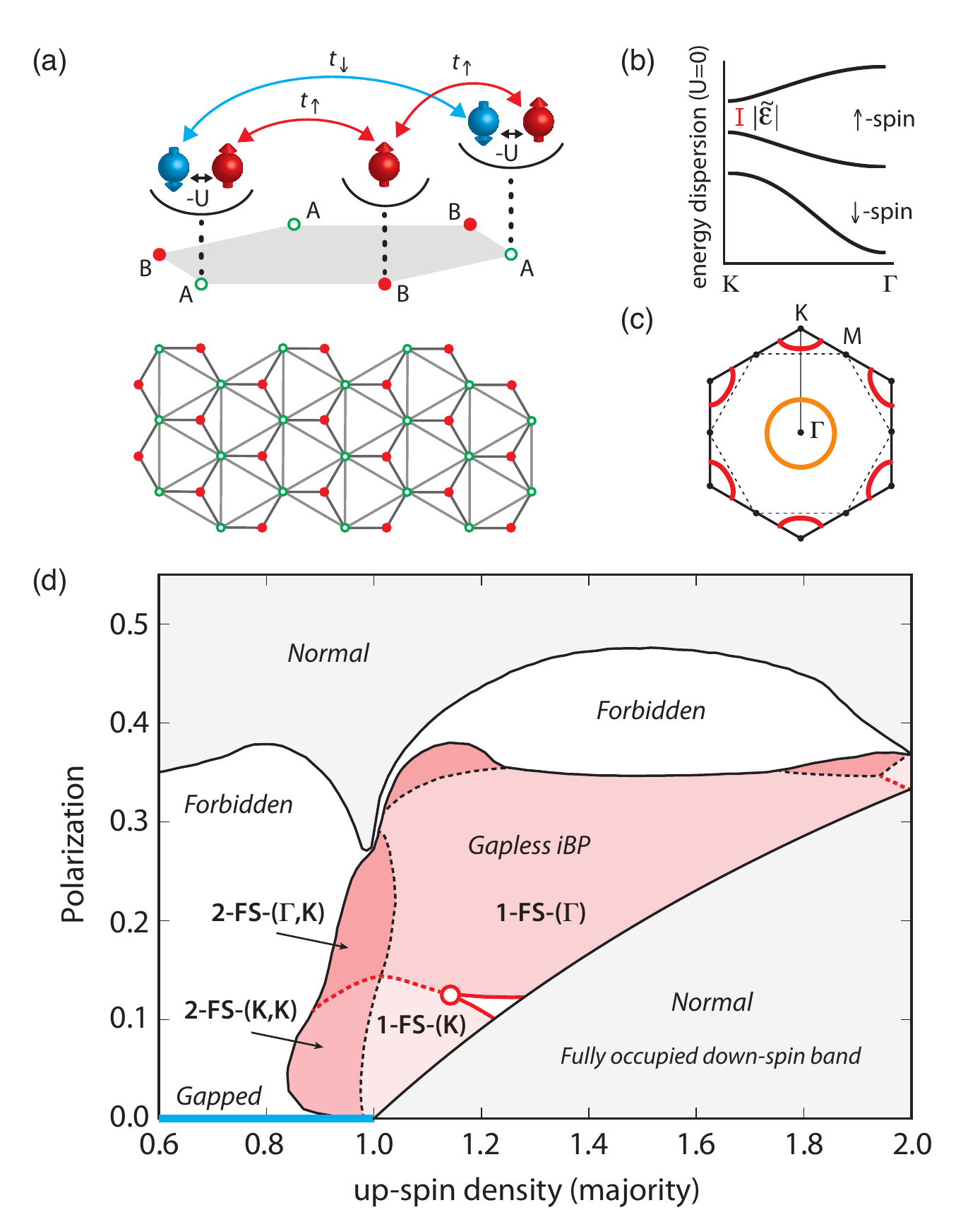}
\caption{
\label{fig1} 
The mixed-geometry lattice proposed and the phase diagram of exotic paired states 
at zero temperature. 
(a) Up- and down-spin components are selectively loaded on honeycomb
(up-spin, $A$ and $B$) 
and triangular (down-spin, $A$) lattices. 
(b) Noninteracting energy dispersions are sketched along the $K-\Gamma$ 
line in the first Brillouin zone shown in (c). The gap in the up-spin bands is tuned 
by the on-site energy modulation $\tilde{\epsilon}$.
(d) The phase diagram is plotted as a function of  the up-spin density $n_\uparrow$ 
and the polarization $P=\frac{n_\uparrow-n_\downarrow}{n_\uparrow+n_\downarrow}$
for the case of $\tilde{\epsilon}=0$.
The gapless states are characterized by the Fermi surface topology
denoted as $z\text{-FS-}(X)$ indicating $z$ Fermi surfaces being 
centered at $X\equiv\Gamma$ and/or $\mathrm{K}$ as is illustrated in (c). 
Continuous (dotted lines) and discontinuous (solid lines) Lifshitz 
transitions are found, and the empty circle indicates the multicritical point.
}
\end{figure}

Ultracold Fermi gases are potentially an ideal test bed of quantum many-body 
physics because of their unprecedented controllability~\cite{Bloch2008}. 
Loaded in optical lattices, fermionic atoms with tunable interparticle interaction 
can construct the Fermi-Hubbard model~\cite{Jordens2008,Schneider2008}.
One remarkable feature of ultracold gases is the tunable
asymmetry between the (pseudo)spins associated with atomic internal states. 
Typical forms of spin asymmetry are population~\cite{MIT,RICE,ENS,Liao2010} 
or mass~\cite{Wille2008,Voigt2009,Kohstall2012} imbalance between the spin components
or spin-dependent optical lattices 
realized for bosons~\cite{LeBlanc2007,Catani2009,Lamporesi2010,SoltanPanahi2011,SoltanPanahi2012}. 
The effects of anisotropic hopping or on-site potentials 
in a lattice~\cite{Feiguin2009,Zapata2010}, as well as
mixed-dimensional pairing in continuum~\cite{Nishida2008,Iskin2010}, 
have been theoretically investigated. 
In this Letter, we consider a fundamentally different type of asymmetry:
a spin-selective lattice where each spin component is separately loaded 
on a different underlying geometry. 
The up- and down-spin components are selectively loaded on honeycomb lattices 
with two bands and triangular lattices with one band, respectively, as shown in Fig.~\ref{fig1}(a). 
This is arguably one of the simplest yet most nontrivial mixed-geometry 
configurations that could be experimentally approached by generalizing the techniques
in \cite{LeBlanc2007,Catani2009,Lamporesi2010,SoltanPanahi2011,SoltanPanahi2012}
or using two different atomic species \cite{Wille2008,Voigt2009,Kohstall2012},
providing a novel multiband pairing scenario.
In addition, the mixed-geometry system that we propose reveals 
a notable structural similarity to hybrid graphene suggested for 
superconductivity~\cite{Uchoa2007,Profeta2012}.

We consider the attractive Hubbard model for a fermion pairing problem                                                                                                                                                                                                                                  
in the mixed-geometry lattices. The system is a superlattice of two 
spin-dependent sublattices $A$ and $B$ [see Fig.~\ref{fig1}(a)].
Turned on short-ranged interactions, this configuration may realize
the Hubbard model with the on-site interaction selectively applied 
at $A$ sites. The Hamiltonian can be written as 
\begin{eqnarray*}
\mathcal{H} = &-& t_\uparrow \sum_{\langle i,j \rangle \in \mathcal{L}_\uparrow}
( \hat{a}^\dagger_{i\uparrow} \hat{b}_{j\uparrow} + \mathrm{H.c.})
+ \epsilon^a_\uparrow \sum_i \hat{n}^a_{i\uparrow}
+ \epsilon^b_\uparrow \sum_i \hat{n}^b_{i\uparrow}
\\
&-& t_\downarrow \sum_{\langle i,j \rangle \in \mathcal{L}_\downarrow} 
( \hat{a}^\dagger_{i\downarrow} \hat{a}_{j\downarrow} + \mathrm{H.c.}) 
+ \epsilon^a_\downarrow \sum_i \hat{n}^a_{i\downarrow} \\
&-& \mu_\uparrow \sum_{i,j} (\hat{n}^a_{i\uparrow} + \hat{n}^b_{j\uparrow})
- \mu_\downarrow \sum_i \hat{n}^a_{i\downarrow}
- U \sum_i \hat{n}^a_{i\uparrow}\hat{n}^a_{i\downarrow}  ~,
\end{eqnarray*}
where $\hat{a}^\dagger$ ($\hat{a}$) and $\hat{b}^\dagger$ ($\hat{b}$) are 
fermionic creation (annihilation) operators in the sublattices $A$ and $B$, 
respectively, and $\hat{n}^a$ and $\hat{n}^b$ 
are corresponding density operators. 
The spin-dependent hopping occurs between nearest-neighboring sites 
$\langle i, j\rangle$ in the up-spin honeycomb lattice $\mathcal{L}_\uparrow$ 
and the down-spin triangular lattice $\mathcal{L}_\downarrow$, while
the hopping strengths $t_\uparrow$ and $t_\downarrow$
are set to be unity for simplicity. Introducing the control parameter $\tilde{\epsilon}$
for on-site energy modulations, the on-site terms are chosen as
$\epsilon^a_\uparrow = -\tilde{\epsilon}/2$,
$\epsilon^b_\uparrow = \tilde{\epsilon}/2$, and 
$\epsilon^a_\downarrow = -3$ without loss of generality.
The on-site interaction is chosen as $U=5$. 

\begin{figure*}
\includegraphics[width=0.95\textwidth]{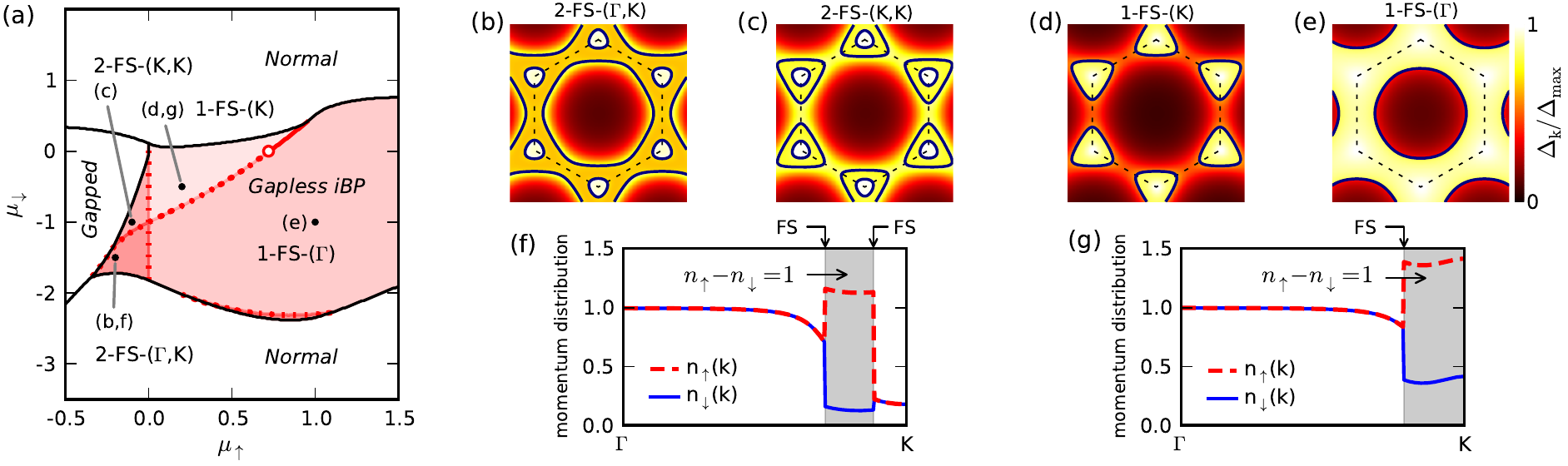}
\caption{
\label{fig2}
Characterization of paired states in the gapless phase. Topologically distinct 
Fermi surface configurations are shown with momentum-resolved order parameter 
$\Delta_k \equiv \langle \tilde{a}_{\text{-}\vec{k}\downarrow} \tilde{a}_{\vec{k}\uparrow}\rangle$ 
in the first Brillouin zone (dashed line) for the cases (b)-(e) 
sampled from the phase diagram (a) for $\tilde{\epsilon}=0$. 
Momentum distributions further characterize the gapless paired states 
with (f) $2$-FS and (g) $1$-FS.
The phase diagram in (a) presents the continuous (dotted lines) and discontinuous (solid lines)
Lifshitz transitions as a function of chemical potentials
$\mu_\uparrow$ and $\mu_\downarrow$, where the multicritical point (empty circle) 
at $\mu_\downarrow=0$ corresponds to a step-function-like singularity 
in the density of states appearing when a noninteracting down-spin band is fully filled.
}
\end{figure*}

The mean-field order parameter is defined as 
$\Delta = U\langle \hat{a}_{i\downarrow} \hat{a}_{i\uparrow} \rangle$.
Neglecting fluctuations, the interaction term can be approximated as 
$-U\sum_i \hat{n}^a_{i\uparrow}\hat{n}^a_{i\downarrow} \approx 
-\sum_i (\Delta \hat{a}^\dagger_{i\uparrow}\hat{a}^\dagger_{i\downarrow}
+ \Delta^* \hat{a}_{i\downarrow}\hat{a}_{i\uparrow}) + |\Delta|^2/U$.  
Performing the Fourier transformation, 
$\tilde{f}_{\vec{k}\sigma} = \sum_i e^{-i\vec{k}\cdot\vec{x}_i} \hat{f}_{i\sigma}$
where $f$ is $a$ or $b$, the mean-field Hamiltonian can be written in momentum space as 
\begin{eqnarray}
\label{eq:Hk}
\mathcal{H}_\mathrm{MF} &=& \sum_{\vec{k}}\sum_{\alpha \in \{1,2,3\}} \xi_\alpha(\vec{k})
\hat{c}_{\alpha \vec{k}}^\dagger \hat{c}_{\alpha \vec{k}} \\
&+& \sum_{\vec{k}} [ g_1(\vec{k})\hat{c}_{1\vec{k}}^\dagger \hat{c}_{3, \text{-}\vec{k}}^\dagger 
+ g_2(\vec{k})\hat{c}_{2\vec{k}}^\dagger \hat{c}_{3, \text{-}\vec{k}}^\dagger + \mathrm{H.c.}] 
\nonumber
\end{eqnarray}
in the basis of noninteracting bands~\cite{supp}, 
providing a \textit{multiband} system where up-spin particles in band $\alpha=1,2$ 
would pair with down-spin particles in band $\alpha=3$.
The lattice effects of the mixed geometry are encoded in the noninteracting 
band dispersion $\xi_\alpha(\vec{k})$ sketched in Fig.~\ref{fig1}(b) 
and in the coupling $g_{1,2}(\vec{k})$. The dispersions are explicitly written as 
$\xi_{1,2}(\vec{k}) = \pm [\tilde{\epsilon}^2 /4+ |h_\uparrow(\vec{k})|^2]^{1/2} - \mu_\uparrow$,
where $h_\uparrow(\vec{k}) = -t_\uparrow [e^\frac{ik_x}{\sqrt{3}} 
+ 2e^\frac{-ik_x}{2\sqrt{3}} \cos \frac{k_y}{2}]$, and
$\xi_3(\vec{k}) = -t_\downarrow [ 2(\cos k_y + \cos \frac{1}{2}(k_y + \sqrt{3}k_x) 
+ \cos \frac{1}{2}(k_y - \sqrt{3}k_x)) + 3] - \mu_\downarrow$.
The interband couplings are derived as
$g_{1,2}(\vec{k}) = - \frac{\Delta}{\sqrt{2}} 
\sqrt{1 \mp \tilde{\epsilon} / (\tilde{\epsilon}^2 + 4|h_\uparrow(\vec{k})|^2)^{1/2} }$.
With fixed interaction and hopping strengths, 
two control parameters can govern the pairing in this system:
(1) tuning chemical potentials can lead to a Fermi surface mismatch, and
(2) the on-site energy modulation $\tilde{\epsilon}$ controls a relative strength 
of the couplings $g_1$ and $g_2$. 
The quasiparticle energies $E_\alpha(\vec{k})$
can be obtained by solving Eq.~(\ref{eq:Hk})~\cite{supp}, and 
the grand potential $\Omega \equiv -\frac{1}{\beta} 
\ln \mathrm{Tr} \exp[-\beta\mathcal{H}_\mathrm{MF}]$ is calculated as 
$\Omega(\Delta) = \frac{|\Delta|^2}{U} + \sum_{\vec{k}} \xi_3(-\vec{k})
+ \frac{1}{\beta} \sum_{\vec{k},\alpha} \ln (1+e^{-\beta E_\alpha(\vec{k})})$.
The ground state is self-consistently obtained by the order parameter $\Delta$ minimizing 
the grand potential $\Omega$. We have confirmed that the global minimum of $\Omega$ is found.

Figure~\ref{fig1}(d) shows the zero-temperature phase diagram for 
the case of the graphenelike up-spin bands ($\tilde{\epsilon}=0$). 
We find that the diagram is divided into
three main areas indicating the normal, gapped, and gapless phases. 
While the normal phase is simply indicated by a vanishing order parameter $\Delta = 0$, 
a paired state with nonzero $\Delta$ is discriminated by its single-particle 
excitation spectrum to be in the gapped or gapless phases.
The gapped phase is a fully paired state, similar to the BCS state.
In the gapless phases, we find an exotic paired state that we call 
the \textit{incomplete-breached-pair} (iBP) state 
because of a \textit{partial} breach found in momentum distribution. 

Intriguingly, the iBP state can be further divided into different subclasses 
by the topological arrangement of one or two Fermi surfaces,
implying that Lifshitz transitions~\cite{Lifshitz1960} occur in the gapless phase.
The iBP states undergo two types of transitions 
where the Fermi surface is either vanishing (2-FS to 1-FS)
or transforming to a topologically different surface 
($\Gamma$ centered to $K$ centered). In this mixed-geometry system,
while the former is a continuous transition, it turns out that
the latter can be continuous or discontinuous transitions 
that meet at a multicritical point [see Figs.~\ref{fig1}(d) and \ref{fig2}(a)]. 
This end point of the discontinuous transition may be explained 
as the marginal quantum critical point of a Lifshitz transition 
in a two-dimensional interacting system~\cite{Yamaji2006}. 

The nature of the iBP states can be characterized by momentum 
distributions $n_{\uparrow,\downarrow}(\vec{k})$ and 
a momentum-resolved order parameter $\Delta_k$ shown in Fig.~\ref{fig2}.
While the jumps in $n_{\uparrow,\downarrow}(\vec{k})$ and $\Delta_k$ indicate 
the presence of the Fermi surfaces, the breach surrounded by these discontinuities 
shows a clear contrast to that of the conventional BP state in momentum space.
It turns out that the momentum-space breach in our iBP state 
shows a mixture of paired particles with finite $\Delta_k$ coexisting with 
unpaired excess majority species, rather than the perfect phase separation 
suggested by the conventional BP state. However, Luttinger's theorem still 
holds for a conserved quantity (here $n_\uparrow-n_\downarrow$) 
in the broken symmetry phase~\cite{Sachdev2006},
explaining the plateau of $n_\uparrow - n_\downarrow =1$ observed between the Fermi surfaces.

Controlled by chemical potentials or densities, topologically different iBP states 
arise with distinct Fermi surface configurations. 
It is not only that the number of the Fermi surfaces 
essentially changes between one and two, as observed, for instance, at $\mu_\uparrow=0$,
but also the shape of the Fermi surface undergoes a topological transition (see Fig.~\ref{fig2}).
Note that the transitions in our phase diagram are driven by densities or 
chemical potentials, which is in contrast to the BEC-BCS crossover 
for imbalanced two-component gases 
where the transition into the BP state with one Fermi surface may occur in the 
BEC regime driven by interaction~\cite{Pao2006,Sheehy2006}.
In addition, compared with a noninteracting case where a topological change of 
the Fermi surface occurs simply at the van Hove singularity 
at $|\mu|=1$ [see Fig.~\ref{fig1}(c)], 
it is notable that the transition lines in the interacting system are very different 
from $|\mu|=1$ as indicated by our diagram 
of the phases with $\Gamma$- and $K$-centered Fermi surfaces. 

\begin{figure}[b]
\includegraphics[width=0.40\textwidth]{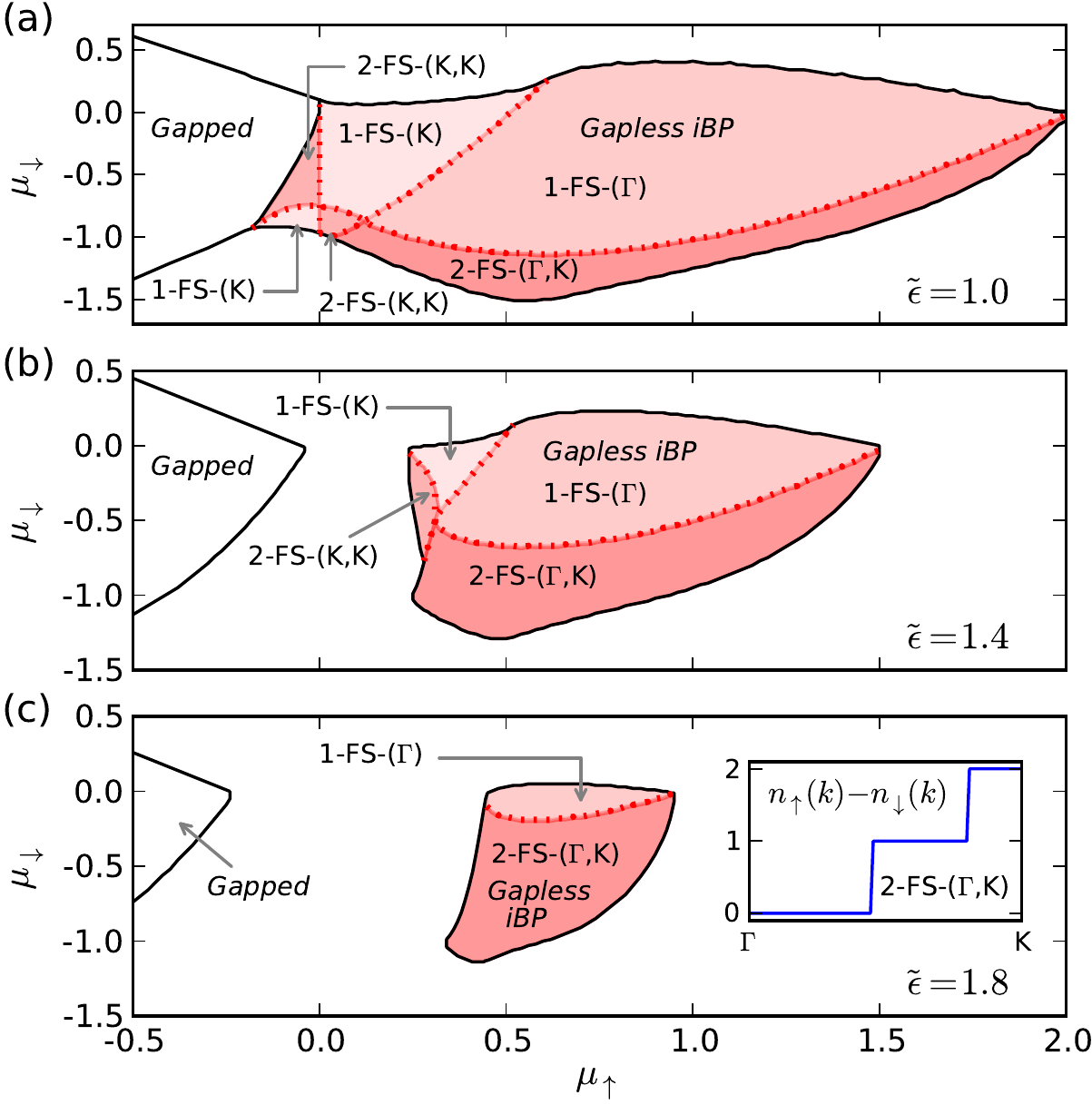}
\caption{
\label{fig3}
Effects of the on-site energy modulation $\tilde{\epsilon}$. The phase diagrams are examined 
for $\tilde{\epsilon}=$ (a) $1.0$, (b) $1.4$, and (c) $1.8$.
While the gapped phase is insensitive to $\tilde{\epsilon}$, the gapless phase (shaded)
shows different Fermi surface topologies with finite $\tilde{\epsilon}$.
In particular, a dominant phase at a large $\tilde{\epsilon}$ 
is found to be the $2\text{-FS-}(\Gamma,K)$ state, which is distinguished from 
the one at $\tilde{\epsilon}=0$ by
its different momentum distribution with a fully polarized area, 
as is illustrated in the inset of (c). 
}
\end{figure}

The characteristics of the gapless phase can also be tuned
by on-site energy modulation $\tilde{\epsilon}$. The effects 
of finite $\tilde{\epsilon}$ are twofold. First, it opens 
a band gap in the noninteracting up-spin bands. Second, it creates 
an imbalance between the interband couplings $g_1$ and $g_2$
[see Eq.~(\ref{eq:Hk})].
Consequently, for a large $\tilde{\epsilon}$ giving $g_2 \gg g_1$, 
the contribution of the up-spin upper band ($\alpha=1$) to the pairing 
can be significantly suppressed.
Figure~\ref{fig3} displays the phase diagrams for $\tilde{\epsilon} = 1.0,1.4,1.8$,
indicating that the gapless phase becomes less robust as $\tilde{\epsilon}$ increases. 
This apparent connection between the coupling imbalance 
and the stability of the gapless phase implies that a well-balanced 
contribution of both interband pairings may be an important factor for our mixed-geometry lattice 
in stabilizing the iBP phase.

\begin{figure}
\includegraphics[width=0.47\textwidth]{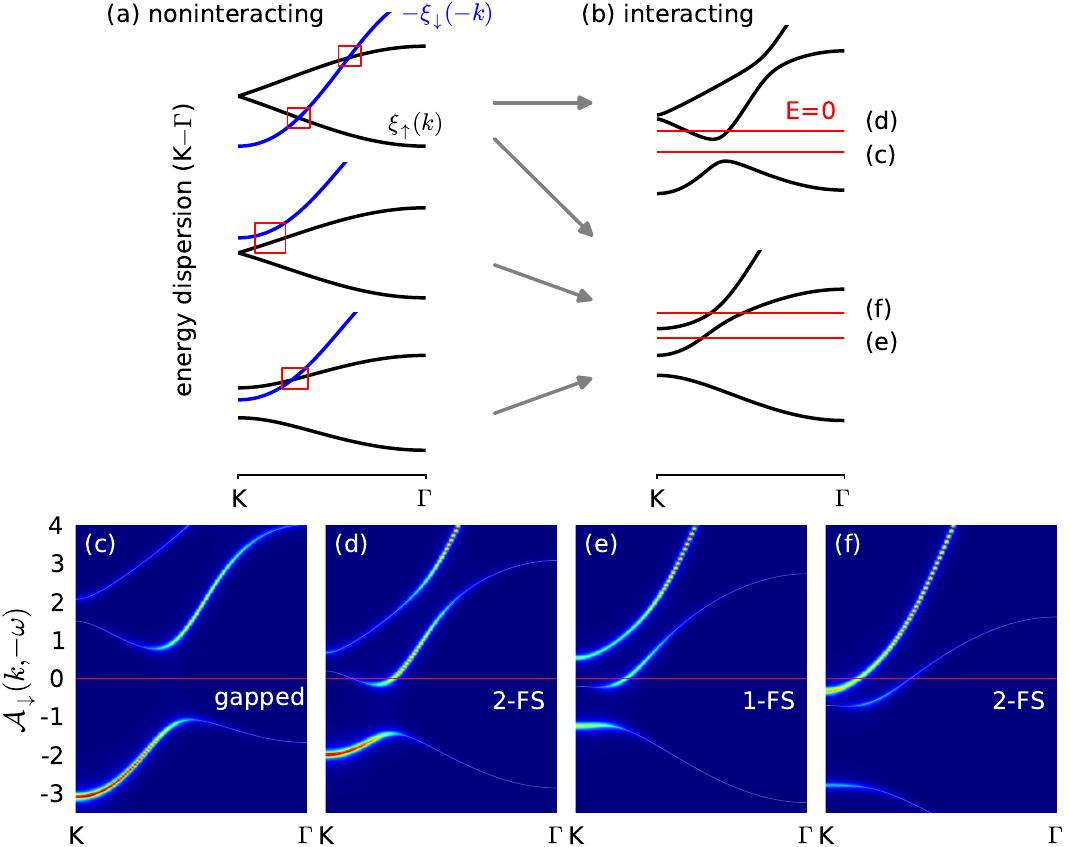}
\caption{
\label{fig4}
Schematics of pairing gap opening and Fermi surface formation. 
The diagrams are plotted along the $K-\Gamma$ line [see Fig.~\ref{fig1}(c)].
The noninteracting particlelike up-spin ($\xi_\uparrow$) and 
holelike down-spin ($-\xi_\downarrow$) bands sketched in (a) evolve 
into (b) the quasiparticle bands with interaction turned on. 
The boxes indicate the areas of gap formation.
The Fermi surface formation depends on the position of the zero-excitation level
($E=0$), as shown in the spectral functions of the down-spin component
$\mathcal{A}_\downarrow (k,-\omega)$ for all typical cases 
with (c) zero, (e) one, and (d), (f) two Fermi surfaces. 
}
\end{figure}

One of our main findings is the chemical-potential-driven emergence of 
the paired states with different Fermi surfaces. The importance 
of the multiband contribution to the formation of these exotic paired states 
can be intuitively understood in the band structure of the mixed-geometry system.
Revisiting our mean-field procedures~\cite{supp}, 
as illustrated in Fig.~\ref{fig4}, there are three cases of zero, one, and two crossings 
allowed between the noninteracting particlelike ($\xi_\uparrow$) 
and holelike ($-\xi_\downarrow$) dispersions with varying chemical potentials. 
The interaction ($g_{1,2}$) causes gap openings at crossings, providing 
the two different configurations of quasiparticle energy dispersion 
that form Fermi surfaces with various topology at given chemical potentials.

The simple band picture can also clarify the arrangements of  
the distinct iBP states in the phase diagrams. 
For instance, considering the iBP state with two Fermi surfaces, 
Fig.~\ref{fig4}(b) indicates that there are two multiband configurations forming such states, 
one in the lower band side (d) and the other in the upper band side (f).
In contrast, the state with one Fermi surface  occurs only in the upper band side (e).
This classification in the band picture is consistent with the phase boundaries 
in the phase diagrams. In addition, 
the band picture indicates that the iBP states with two Fermi surfaces have 
a different nature depending on whether they originate from the upper or lower band side. 
Indeed, for instance, the momentum distributions found at $\tilde{\epsilon}=1.8$ 
show a fully polarized normal area of excess majority particles around $K$, 
which is in contrast to its fully paired counterpart 
found at $\tilde{\epsilon}=0$~\cite{supp}.
A systematic study of finite temperature effects and quantum fluctuations 
is beyond the scope of this work. While the Fermi surface is strictly defined
at zero temperature, we have tested that the main features 
are present at finite-temperatures within the mean-field theory, 
critical temperatures typically being of the order of one-tenth of the hopping amplitude.

Mixed-geometry lattice systems such as those we propose here extend the realm of 
pairing and strong correlation physics. The attractive Hubbard model 
in the mixed honeycomb-triangular lattice that we have considered 
reveals a new exotic multiband paired state, the iBP state, 
and shows a rich phase diagram of topologically distinct phases. 
We have found that a degree of the mixing of the two up-spin bands in pairing up 
with the down-spin band plays a key role in stabilizing the exotic paired state. 
Our findings thus contribute to a fundamental understanding of multiband effects, 
in general, which are relevant to many real superconducting materials.
The high critical temperature of MgB$_2$ is likely due to the coupling of order parameters 
in multiple bands~\cite{Suhl1959, Moskalenko1959}, and, in iron pnictides, 
interband mechanisms have been suggested to play an important role. 
Furthermore, the similarity of our mixed-geometry
lattice to the hybrid graphene systems~\cite{Uchoa2007,Profeta2012} 
may inspire studies of exotic superconductivity in such designed nanomaterials. 
Therefore, realizing mixed-geometry optical lattices 
would open a new dimension in the search for novel quantum states using ultracold gases 
as a quantum simulator.

\begin{acknowledgments}
We are grateful to S. Roche for illuminating discussions.
This work was supported by the Academy of Finland through its
Centre of Excellence Programme 
(Projects No. 139514, No. 141039, No. 251748, No. 263347, and No. 135000). 
Computing resources were provided by CSC$-$the Finnish IT Centre for Science and the Aalto Science-IT Project.
\end{acknowledgments}


\begin{thebibliography}{}

\bibitem{FF}
P. Fulde and R. A. Ferrell, Phys. Rev. \textbf{135}, A550 (1964).

\bibitem{LO}
A. I. Larkin and Y. N. Ovchinnikov, Zh. Eksp. Teor. Fiz. \textbf{47}, 1136 (1964) 
[Sov. Phys. JETP \textbf{20}, 762 (1965)].

\bibitem{Sarma}
G. Sarma, J. Phys. Chem. Solids \textbf{24}, 1029 (1963).

\bibitem{BP}
W. V. Liu and F. Wilczek, Phys. Rev. Lett. \textbf{90}, 047002 (2003).

\bibitem{Radovan2003} 
H. A. Radovan, N. A. Fortune, T. P. Murphy, S. T. Hannahs, E. C. Palm,
S. Tozer, and D. Hall, Nature (London) \textbf{425}, 51 (2003).

\bibitem{Kenzelmann2008}
M. Kenzelmann, Th. Str\"assle, C. Niedermayer, M. Sigrist, B. Padmanabhan,
M. Zolliker, A. D. Bianchi, R. Movshovich, E. D. Bauer,
J. L. Sarrao, and J. D. Thompson, Science \textbf{321}, 1652 (2008).

\bibitem{MIT}
M. W. Zwierlein, A. Schirotzek, C. H. Schunck, and W. Ketterle,  
Science \textbf{311}, 492 (2006).

\bibitem{RICE}
G. B. Partridge, W. Li, R. I. Kamar, Y. Liao, and R. G. Hulet,
Science \textbf{311}, 503 (2006).

\bibitem{ENS}
S. Nascimb\'ene, N. Navon, K. J. Jiang, L. Tarruell, M. Teichmann, J. McKeever, F. Chevy, and C. Salomon, 
Phys. Rev. Lett. \textbf{103}, 170402 (2009).

\bibitem{Liao2010}
Y. Liao, A. S. C. Rittner, T. Paprotta, W. Li, G. B. Partridge, R. G. Hulet, S. K. Baur, and E. J. Mueller, 
Nature (London) \textbf{467}, 567 (2010).

\bibitem{Bloch2008}
I. Bloch, J. Dalibard, and W. Zwerger, Rev. Mod. Phys. \textbf{80}, 885 (2008).

\bibitem{Jordens2008}
R. J\"ordens, N. Strohmaier, K. G\"unter, H. Moritz, and T. Esslinger, 
Nature (London) \textbf{455}, 204 (2008).

\bibitem{Schneider2008}
U. Schneider, L. Hackerm\"uller, S. Will, Th. Best, I. Bloch, 
T. A. Costi, R. W. Helmes, D. Rasch, and A. Rosch, 
Science \textbf{322}, 1520 (2008).

\bibitem{Wille2008}
E. Wille, F. M. Spiegelhalder, G. Kerner, D. Naik, A. Trenkwalder,
G. Hendl, F. Schreck, R. Grimm, T. G. Tiecke, J. T. M. Walraven,
S. J. J. M. F. Kokkelmans, E. Tiesinga, and P. S. Julienne,
Phys. Rev. Lett. \textbf{100}, 053201 (2008).

\bibitem{Voigt2009}
A.-C. Voigt, M. Taglieber, L. Costa, T. Aoki, W. Wieser, T. W. H\"ansch, and K. Dieckmann,
Phys. Rev. Lett. \textbf{102}, 020405 (2009).

\bibitem{Kohstall2012}
C. Kohstall, M. Zaccanti, M. Jag, A. Trenkwalder, P. Massignan,	G. M. Bruun, 
F. Schreck, and R. Grimm, Nature (London) \textbf{485}, 615 (2012).

\bibitem{LeBlanc2007}
L. J. LeBlanc and J. H. Thywissen, Phys. Rev. A \textbf{75}, 053612 (2007).

\bibitem{Catani2009}
J. Catani,  G. Barontini, G. Lamporesi, F. Rabatti, G. Thalhammer, F. Minardi, S. Stringari, and M. Inguscio,
Phys. Rev. Lett. \textbf{103}, 140401 (2009).

\bibitem{Lamporesi2010}
G. Lamporesi, J. Catani, G. Barontini, Y. Nishida, M. Inguscio, and F. Minardi,
Phys. Rev. Lett. \textbf{104}, 153202 (2010).

\bibitem{SoltanPanahi2011}
P. Soltan-Panahi, J. Struck, P. Hauke, A. Bick, W. Plenkers, G. Meineke, C. Becker, P. Windpassinger, M. Lewenstein, and K. Sengstock, 
Nat. Phys. \textbf{7}, 434 (2011).

\bibitem{SoltanPanahi2012}
P. Soltan-Panahi, D.-S. L\"uhmann, J. Struck, P. Windpassinger, and K. Sengstock, 
Nat. Phys. \textbf{8}, 71 (2012).

\bibitem{Feiguin2009}
A. E. Feiguin and M. P. A. Fisher, Phys. Rev. Lett. \textbf{103}, 025303 (2009).

\bibitem{Zapata2010}
I. Zapata, B. Wunsch, N. T. Zinner, and E. Demler,
Phys. Rev. Lett. \textbf{105}, 095301 (2010).

\bibitem{Nishida2008}
Y. Nishida and S. Tan, Phys. Rev. Lett. \textbf{101}, 170401 (2008).

\bibitem{Iskin2010}
M. Iskin and A. L. Suba\c{s}i, Phys. Rev. A \textbf{82}, 063628 (2010).

\bibitem{Uchoa2007}
B. Uchoa and A. H. Castro Neto, Phys. Rev. Lett. \textbf{98}, 146801 (2007).

\bibitem{Profeta2012}
G. Profeta, M. Calandra, and F. Mauri, Nature Phys. \textbf{8}, 131 (2012).

\bibitem{supp}
See Supplemental Material for the detailed derivation and further numerical results.

\bibitem{Lifshitz1960}
I. M. Lifshitz, Zh. Eksp. Teor. Fiz. \textbf{38}, 1569 (1960)
[Sov. Phys. JETP \textbf{11}, 1130 (1960)]. 

\bibitem{Yamaji2006}
Y. Yamaji, T. Misawa, and M. Imada, J. Phys. Soc. Jpn \textbf{75}, 094719 (2006).

\bibitem{Sachdev2006}
S. Sachdev and K. Yang, Phys. Rev. B \textbf{73}, 174504 (2006). 

\bibitem{Pao2006}
C.-H. Pao, S.-T. Wu, and S.-K. Yip, Phys. Rev. B \textbf{73}, 132506 (2006).

\bibitem{Sheehy2006}
D. E. Sheehy and L. Radzihovsky, Phys. Rev. Lett. \textbf{96}, 060401 (2006).

\bibitem{Suhl1959}
H. Suhl, B. T. Matthias, and L. R. Walker, Phys. Rev. Lett. \textbf{3}, 552 (1959).

\bibitem{Moskalenko1959}
V. A. Moskalenko, Fiz. Met. Metalloved. \textbf{8}, 503 (1959)
[Phys. Met. Metallogr. (USSR) \textbf{8}, 25 (1959)].


\end{thebibliography}
\end{document}